\documentclass[%
 reprint,
 amsmath,amssymb,
 aps,
]{revtex4-2}

\usepackage{graphicx}
\usepackage{dcolumn}
\usepackage{bm}
\usepackage{xcolor}

\begin{document}

\title{Optical control of collective  states in 1D ordered atomic chains beyond the linear regime}
\author{N. Fayard}\email[Corresponding author: ]{nikos.fayard@institutoptique.fr}
\address{Universit{\'e} Paris-Saclay, Institut d'Optique Graduate School, CNRS, Laboratoire Charles Fabry, 91127
Palaiseau, France}
\author{I. Ferrier-Barbut}
\address{Universit{\'e} Paris-Saclay, Institut d'Optique Graduate School, CNRS, Laboratoire Charles Fabry, 91127
Palaiseau, France}
\author{A. Browaeys}
\address{Universit{\'e} Paris-Saclay, Institut d'Optique Graduate School, CNRS, Laboratoire Charles Fabry, 91127
Palaiseau, France}
\author{J.-J. Greffet}
\address{Universit{\'e} Paris-Saclay, Institut d'Optique Graduate School, CNRS, Laboratoire Charles Fabry, 91127
Palaiseau, France}

\date{\today}
\begin{abstract}
Driven by the need to develop efficient atom-photon interfaces, recent efforts have proposed replacing cavities by large arrays of cold atoms that can support subradiant or superradiant collective states.  In practice, subradiant states are decoupled from radiation, which constitutes  a hurdle to most applications.
In this work, we study theoretically a protocol that bypasses this limit using a one dimensional ($1$D) chain composed of $N$ three-level atoms in a V-shaped configuration. Throughout the protocol, the chain behaves as a time-varying metamaterial: enabling absorption, storage and on-demand emission in a spectrally and spatially controlled  mode. 
Taking into account the quantum nature of atoms, we establish the boundary between the linear regime and the nonlinear regime. In the nonlinear regime, we  demonstrate that doubly-excited states can be coherently  transferred from  superradiant to subradiant states, opening the way to the optical characterization of their entanglement.
\end{abstract}

\maketitle

\section{Introduction}
The atoms' ability to store and process quantum information combined with the minimal loss through light propagation make atom-photon platforms prime candidates for realizing scalable quantum networks~\cite{hammerer2010quantum,raimond2001manipulating,saffman2010quantum}.
In practice,  it is necessary to enhance the light-matter interaction in order to enable  the transfer of the information between the atoms and the photons~\cite{kimble2008quantum}.
To this aim, one possibility consists in interfacing  atoms to cavities~\cite{thompson2013coupling,reiserer2015cavity,plankensteiner2017cavity,shlesinger2021time,lei2023many}, 
 nanofibers~\cite{nayak2007optical,solano2017super,corzo2019waveguide,pennetta2022collective} or  photonic crystal waveguides~\cite{goban2014atom,goban2015superradiance,lodahl2015interfacing,yu2014nanowire,Fayard22,bouscal2023systematic}  which results in quasi one dimensional spontaneous emission~\cite{turchette1995measurement}  and increased  light-matter interaction. Those features triggered a variety of theoretical
proposals for applications in quantum optics and many-body physics~\cite{chang2014quantum,roy2017colloquium}. As of today, most of their experimental realizations have remained elusive  in the many-body regime  due the difficulty of interfacing efficiently a large number of atoms with nanostructures~\cite{sheremet2023waveguide}.

\begin{figure*}   
\begin{center}     
\includegraphics[width=0.8\linewidth]{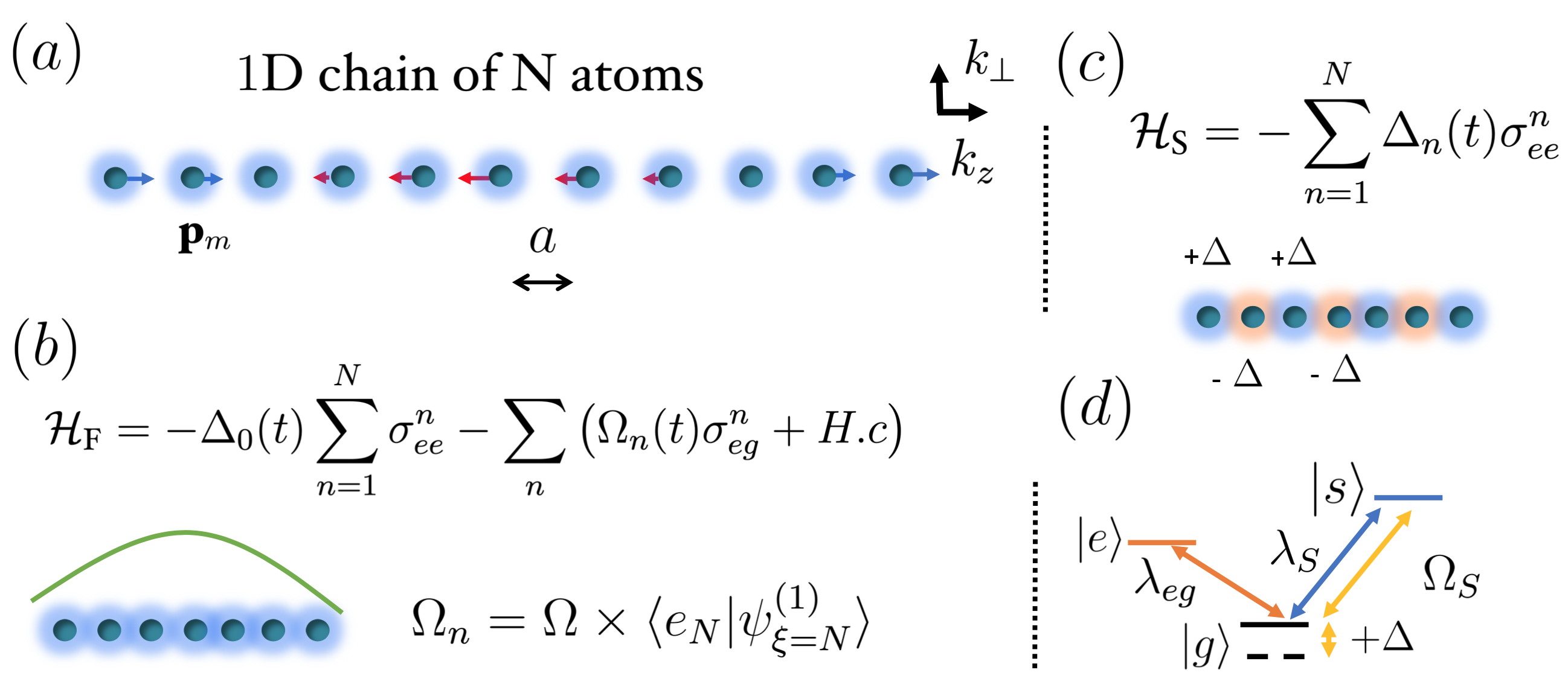} 
\end{center}
   \caption{a) Scheme of a $1$D atomic chain  composed of $N$ atoms at positions $\mathbf{r}_m=(0,0,ma)$ and represented with a polarization linear and parallel to the chain $\mathbf{p}_m=(0,0,p_m)$. (b) Description of the excitation part of the protocol. We illuminate transversally the chain with a field whose spatial profile $\Omega_n$ and frequency $\omega_L$ are adjusted to excite the most superradiant singly-excited eigenmode $\vert \psi _{\xi=N} ^{(1)}\rangle$ .  (c) Description of the transfer part of the scheme. We use a far off resonant electric field to create a  detuning pattern whose sign can vary from an atom to its neighbour. (d) Representation of a V-shaped transition of an atom. The application of a far off resonant electric field $\Omega_S$ with respect to the $\vert g\rangle-\vert s\rangle$ transition  allows to manipulate the detuning of the  $\vert g\rangle-\vert e\rangle$  transition   with a spatial resolution given by $\lambda_S$ potentially smaller than $\lambda_{eg}$. In the case of Dysprosium, $\lambda_{eg}=741$nm and $\lambda_{S}=421$nm.}
 \label{fig:schema}   
\end{figure*}

Recently, it has been shown that  strong light-matter interaction can also emerge in large arrays of atoms assembled 
 in free-space~\cite{bettles2016enhanced,facchinetti2016storing,shahmoon2017cooperative,rui2020subradiant}.
 Indeed, when the interatomic distance is reduced below the wavelength, the  light-induced interaction between atoms becomes strong enough to alter the modes of the  atomic array which behaves as a metasurface.  As a result,  some of the modes become superradiant~\cite{dicke1954coherence,gross1982superradiance,scully2006directed,araujo2016superradiance,he2020atomic,he2020geometric,rastogi2022superradiance} and are associated with an enhanced radiation rate while others become subradiant~\cite{scully2015single,plankensteiner2015selective,guerin2016subradiance,cipris2021subradiance,ferioli2021storage} and are decoupled from the radiation.
From an experimental point of view, the ability to create large arrays of neutral atoms with $\mu m$ interatomic distances has already been demonstrated in order to perform quantum simulations ~\cite{bloch2012quantum,nogrette2014single,endres2016atom,barredo2016atom}. However,  the reduction of the atomic distance below the wavelength is  extremely challenging~\cite{rui2020subradiant,srakaew2023subwavelength}. For this purpose, we consider a platform composed of $3$-levels atoms in a V-shaped configuration.
 The role of the second transition is dual. First, it  enables the optical trapping of the atoms  with interatomic distances sufficiently small such that subradiance emerges. Second, it permits   the local manipulation of the atomic transition  frequencies  using light shifts and  address the subradiant modes of the array~\cite{rui2020subradiant,cipris2021subradiance} as proposed recently~\cite{ballantine2021quantum,rubies2022photon}. So far, those theoretical studies have considered the interaction of atoms with single photons so that a classical scattering description is valid. 
Motivated by the possibility to implement this scheme in ongoing experiments using Sr, Yb and Dy~\cite{norcia2018microscopic,saskin2019narrow}, we study theoretically a similar protocol  for a $1$D atomic chain and go beyond the linear regime.
This enables the exploration of  the onset of the many-body regime,  where non-linear effects appear. It also permits the description of the illumination of the chain by an intense coherent pulse. Taking advantage of these possibilities,  we present a method to: 
(i) excite the most superradiant singly-excited state of the chain from the far field with an intense coherent pulse,
(ii) transfer the excitation to the most subradiant state using a position varying detuning (PVD), 
(iii) store the excitation,
and (iv)  emit the photon in the \emph{desired} superradiant mode of the chain, thus controlling  its frequency, its emission rate and pattern.

Our theoretical analysis is valid  beyond the linear regime and  reveals that the doubly-excited population follows a similar trajectory than the singly-excited one.
In particular, we demonstrate that we can transfer the two-photon population at will  into its most subradiant form where it can be stored and studied.

\section{Description of the system}
In this work, we explore a simple and robust physical system that could be built in a lab capable of absorbing, storing and emitting a photon with a temporal and spatial control. It consists in a linear chain of atoms with a total length larger than the wavelength to enable directivity and with a spacing smaller than half a wavelength to enhance interactions  as represented in Fig.~\ref{fig:schema}(a). It is composed of $N$ three-level atoms ($\vert g\rangle,\vert e\rangle,\vert s\rangle$) located at positions $\mathbf{r}_m=(0,0,m a)$. The frequency of the   $\vert g\rangle-\vert e\rangle$  transition is given by $\omega_{eg}=2\pi c/\lambda_{eg}$ and its polarization is considered linear 
and parallel to the chain.
A second transition depicted in Fig.~\ref{fig:schema}(d)  exists between $\vert g\rangle$ and $\vert s\rangle$ with a transition frequency $\omega_S=2 \pi/\lambda_S$. The role of this transition is dual.  First, it enables the optical trapping of the atoms with a nearest neighbor distance $a< \lambda_{eg}/2$ sufficiently small for subradiant modes to emerge.
Second, it enables the realization of arbitrary detuning  patterns~\cite{de2017optical} between the different atoms. It is a key  point since  the method we follow to manipulate collective states~\cite{ballantine2021quantum,rubies2022photon}  necessitate the realization of the so-called staggered pattern for which  the local detuning of atom $n$  changes sign from site to site: $\Delta_n=(-1)^n\Delta$.
As a concrete example, we  work  with a conservative value of $a/\lambda_{eg}=0.35<0.5$ sufficiently small to  create  arbitrary PVD patterns and manipulate the subradiant modes that emerge in the atomic chain~\cite{rubies2022photon,ballantine2021quantum}.  This can be realized experimentally by loading an accordion lattice from an optical tweezer array~\cite{ville2017loading}.
 
\begin{figure*}   
\begin{center}     
\includegraphics[width=0.8\linewidth]{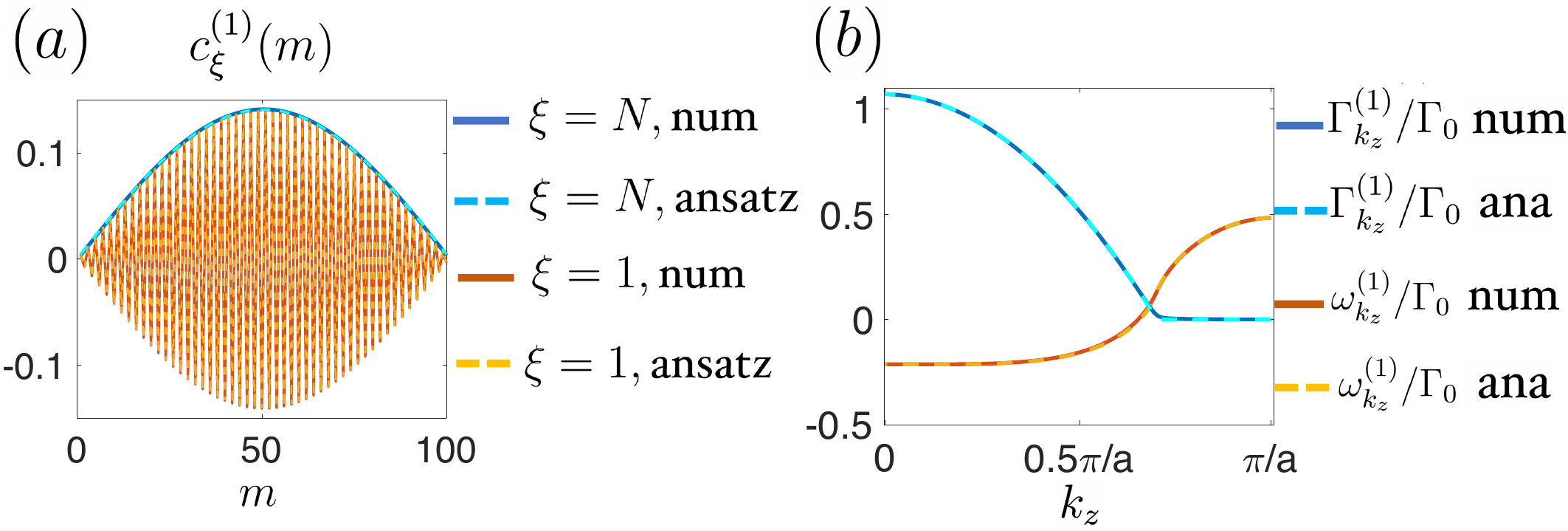} 
\end{center}
   \caption{a) Projection of  $\vert \psi _\xi ^{(1)}\rangle=\sum_{m=1}^Nc_{\xi}^{(1)}(m)\vert e_i\rangle$ on the physical basis of a chain composed of $N=100$ atoms, for both the most  superradiant ($\xi=N$ in blue) and subradiant singly-excited modes  ($\xi=1$ in red). We  compare  with the expression given by the singly-excited ansatz $\vert \psi _\xi ^{\textrm{ans}}\rangle=\sqrt{\frac{2}{N+1}}\sum_{m=1}^N\textrm{sin}\left(\frac{[N+1-\xi] m \pi}{N+1}\right)\vert e_m\rangle$ in dashed. b) Numerical evaluation of $\omega_{k_z}^{(1)}/\Gamma_0$  (solid red) and $\Gamma_{k_z}^{(1)}/\Gamma_0$ (solid blue)  together with their analytical expression given in Eq.~\ref{eq:anagamma} and Eq.~\ref{eq:anaomega}. The parameters are $N=100$ and $a=0.35\lambda_{eg}$ and the atomic polarization is linear and parallel to the chain.}
 \label{fig:collective}   
\end{figure*}

\section{Theoretical model}
We now introduce the model used to describe the light chain interaction. 
We use the Born and Markov approximation~\cite{reitz2022cooperative,agarwal2012quantum} to integrate out the photonic degrees of freedom  from the full atom-light
system. This results in an interacting and open spin model, describing the dynamics of the atomic density matrix $\rho$ with the Master equation (M.E):
\begin{equation}
\dot{\rho}=-i\left[\mathcal{H}\rho-\rho\mathcal{H}^\dagger\right]+\sum_{m,n}\Gamma_{m,n}\sigma_{eg}^m\rho\sigma_{ge}^n.
\label{eq:Mastereq}
\end{equation}

In this expression $\mathcal{H}=\mathcal{H}_{\textrm{eff}}+\mathcal{H}_F+\mathcal{H}_S$ is the total Hamiltonian  that includes : i)  the effective Hamiltonian between the atoms $\mathcal{H}_{\textrm{eff}}$ from which collective subradiant and superradiant modes emerge, ii) the interacting Hamiltonian between the atoms and the excitation field  $\mathcal{H}_F$ and iii) the interacting Hamiltonian induced by a detuning pattern $\mathcal{H}_S$. 
Both the  population recycling term $\Gamma_{m,n}=2\mu_0\omega_0^2\textrm{Im}\left[\mathbf{p}_n^*\overline{\mathbf{G}}(\mathbf{r}_n,\mathbf{r}_m,\omega_0)\mathbf{p}_m\right]$  and the effective Hamiltonian $\mathcal{H}_{\textrm{eff}}=-\mu_0\omega_0^2\sum_{m,n=1}^N\mathbf{p}_n^*\overline{\mathbf{G}}(\mathbf{r}_n,\mathbf{r}_m,\omega_0)\mathbf{p}_m\sigma_{eg}^m\sigma_{ge}^n$ 
depend on the free space dyadic electromagnetic Green's function  $\overline{\mathbf{G}}$,  the vacuum permeability $\mu_0$,  the dipole
 of atom $m$:  $\mathbf{p}_m$, and $\sigma_{\alpha \beta}=\vert \alpha\rangle \langle \beta \vert$ with $\{\alpha,\beta\}\in\{g,e\}$.  
We model the excitation of the atoms with an intense coherent pulse using: $\mathcal{H}_{\textrm{F}}=-\Delta_0(t)\sum_{n=1}^N\sigma_{ee}^n-\sum_n\left( \Omega_n(t)\sigma_{eg}^n + \textrm{H.c} \right)$, with  $\Delta_0=\omega_L-\omega_{eg}$  the detuning between the laser and  atomic frequencies and $ \Omega_n$ the  Rabi frequency of atom $n$.    
We represent the application of the  PVD using $\mathcal{H}_{\textrm{S}}=-\sum_{n=1}^N\Delta_n(t)\sigma_{ee}^n$, with $\Delta_n$  the local detuning of atom $n$. 
When needed, we can turn off the excitation and/or the detuning pattern setting $\mathcal{H}_{\textrm{F}}$ and/or $\mathcal{H}_{\textrm{S}}$  to $0$ in Eq.~(\ref{eq:Mastereq}).

\subsection{Spectral properties of singly and doubly-excited states.}
To simplify the analysis of Eq.~(\ref{eq:Mastereq}), we sort the eigenstates of  $\mathcal{H}_{\textrm{eff}}$ into different manifolds with a given number of excitations $n_{\textrm{exc}}=(1, 2, ...,N)$ and  use the superscript $(...)^{(n_{\textrm{exc}})}$ to label a given manifold.   
If for some specific reason, $n_{\textrm{exc}}\leq n_{\textrm{max}}$  during the full dynamics,  we  truncate the Hilbert space to the populated manifolds only. 
Here, most of the results are presented with a numerical  truncation of the Hilbert space  to $n_{\textrm{exc}}\leq 2$.
This both reduces the computation time  and simplifies the  analysis to the first two manifolds that we briefly describe below.

The diagonalization of  $\mathcal{H}_{\textrm{eff}}$ in the first manifold results in $N$ eigenvectors $\vert \psi _\xi ^{(1)}\rangle=\sum_{m=1}^Nc_{\xi}^{(1)}(m)\vert e_m\rangle$ such that $\mathcal{H}_{\textrm{eff}}\vert \psi _\xi ^{(1)}\rangle=(\omega_{\xi}^{(1)}-i\Gamma_{\xi}^{(1)}/2)\vert \psi _\xi ^{(1)}\rangle$ where $\omega_{\xi}^{(1)}$ represents the shift in energy of $\vert \psi _\xi ^{(1)}\rangle$ with respect to $\omega_{eg}$ and $\Gamma_{\xi}^{(1)}$ its decay rate. 
In order to differentiate easily the subradiant from the superradiant eigenmodes,  we sort them from smallest to largest decay rate and label them with an integer denoted  $\xi$. For exemple, $\vert \psi _{\xi=1} ^{(1)}\rangle$ is the most subradiant mode of the chain while $\vert \psi _{\xi=N} ^{(1)}\rangle$ is the most superradiant.
In Fig.~\ref{fig:collective}(a), we plot  the amplitude of  $\vert \psi _\xi ^{(1)}\rangle$ on each atom  for both the most  superradiant ($\xi=N$) and subradiant singly-excited modes  ($\xi=1$). 
We see that the singly-excited eigenmodes of an ordered chain are spin waves, well represented by the  ansatz:  $\vert \psi _\xi ^{\textrm{ans}}\rangle=\sqrt{\frac{2}{N+1}}\sum_{m=1}^N\textrm{sin}\left(\frac{[N+1-\xi] m \pi}{N+1}\right)\vert e_m\rangle$~\cite{shahmoon2017cooperative,asenjo2017exponential}.  Qualitatively, the most subradiant mode corresponds to a $\pi$ phase difference between nearest neighbors whereas the superradiant mode corresponds to a uniform phase across the chain. 
 In Fig.~\ref{fig:collective}(b), we compute the decay rate and the frequency shift of each eigenmode and plot them as a function of  $k_z(\xi)=\frac{\pi(N+1-\xi)}{a(N+1)}$. We observe strong variations of the collective properties as a function of $k_z$, in perfect agreement with the analytical expressions given in  appendix A.
In particular, Fig.~\ref{fig:collective}(b) clarifies the physics of super and subradiance:
spin waves with a wave vector along the chain larger than $\omega_{eg}/c$ cannot couple to electromagnetic waves  in vacuum and are strongly subradiant.  On the opposite, superradiant modes are associated with small values of $k_z<\omega_{eg}/c$~\cite{manzoni2018optimization,zhang2019theory,zhang2020subradiant2}.

\begin{figure*}   
\begin{center}     
\includegraphics[width=1\linewidth]{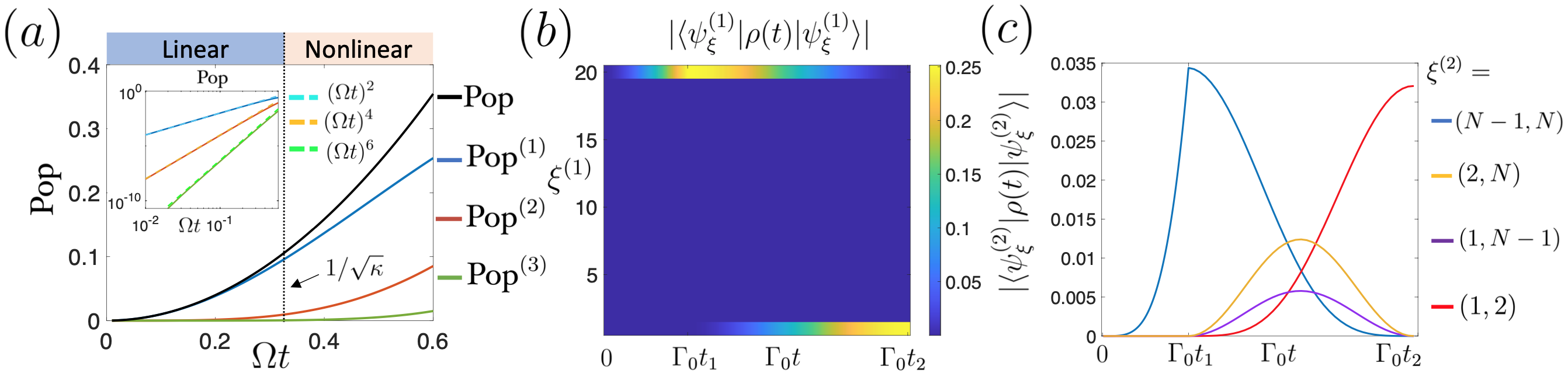} 
\end{center}
   \caption{a) Atomic population as a function of time during the illumination  of the atomic system. The incident monochromatic field is adjusted spatially and spectrally  to couple the ground state of the atomic system to  the most superradiant singly-excited eigenmode. When $\Omega t<1$, the population in each manifold (indicated by different colors) varies like $(\Omega t)^{2n_{\textrm{exc}}}$. The dotted vertical line represents the boundary between the linear and the nonlinear regime that depends on the ratio $\kappa=\Gamma_{\xi=(1,2)}^{(2)}/\Gamma_{\xi=1}^{(1)}$ as discussed in appendix D. b) Projection of the density matrix onto each singly-excited eigenmodes during the first two steps of the protocol as a function of $\Gamma_0 t$. The illumination stops at $t_1=0.6/\Omega$ and the transfer at $t_2=t_1+\pi/(2\Delta)$. c) Projection of the density matrix onto the four doubly-excited eigenmodes that dominates the doubly-excited population  (indicated by different colors) during the first two steps of the protocol. Their label $\xi^{(2)}$  is indicated in the caption.  The numerical parameters are $N=20$, $a=0.35\lambda_{eg}$, the atomic polarization is along the chain and the M.E equation has been truncated to $3$ excitations in (a) and $2$ in (b,c). $\Omega=100\Gamma_0$ and $\Delta=100\Gamma_0$ in (b,c).}
 \label{fig:steps}   
\end{figure*}

We now turn to the excited manifolds and consider states with two excitations.  
 Due to the atomic nonlinearity, highly excited states are in general entangled, and very  different from singly-excited ones~\cite{chang2018colloquium,zhang2020subradiant,bettles2020quantum,moreno2021quantum,fayard2021many,holzinger2022control}. Yet, in the specific case of a $1$D ordered atomic chain composed of $N\gg1$ atoms,  the $N(N-1)/2$ doubly-excited states can be  built as an antisymmetric product of singly-excited spin waves~\cite{asenjo2017exponential,zhang2020subradiant2}. 
 More precisely, we represent  each doubly-excited state using the fermionic ansatz~\cite{asenjo2017exponential}: $\vert \psi _{(\xi_1,\xi_2)} ^{\textrm{ans}}\rangle= \sum_{m<n}\left[c^{(1)}_{\xi_1}(m)c^{(1)}_{\xi_2}(n)-c^{(1)}_{\xi_2}(m)c^{(1)}_{\xi_1}(n)\right]\sigma_{eg}^m\sigma_{eg}^n\vert g\rangle$. In this expression, we used $\xi^{(2)}=(\xi_1,\xi_2)$ to label this state and  note that it is identical to the state labeled with $(\xi_2,\xi_1)$ up to a minus sign.
The consequence of the fermionic ansatz is that each doubly-excited state behaves  like a doubly-excited spin wave, whose collective properties are well approximated by the sum of its single photon components: $\Gamma^{(2)}_{(\xi_1,\xi_2)}\sim \Gamma^{(1)}_{\xi_1}+\Gamma^{(1)}_{\xi_2}$ and $\omega^{(2)}_{(\xi_1,\xi_2)}\sim \omega^{(1)}_{\xi_1}+\omega^{(1)}_{\xi_2}$.  
Let us conclude this section highlighting that the entanglement contained in the atomic degrees of freedom of a highly-excited state is an interesting resource whose extraction remains elusive for now~\cite{masson2020many,bettles2020quantum,cidrim2020photon,williamson2020superatom,moreno2021quantum,zhang2022photon,richter2023collective}.
A protocol, that permits the  transfer of this entanglement to the emitted photons at a given time, and in a given direction would be key for
quantum technologies.

\section{Description of the protocol}
We detail below the different steps of the protocol.
\subsection{Exciting the most superradiant mode.}
We start by illuminating transversally the atomic system from the far field with a strong ($\Omega \gg \Gamma^{(1)}_{\xi=N}$) and monochromatic laser field   resonant with the most superradiant singly-excited eigenmode: $\omega_L=\omega_{eg}+\omega_{\xi=N}^{(1)}$. The spatial profile of the incident field: 
$\Omega_n=\Omega\times\langle e_n\vert \psi_{\xi=N}^{(1)}\rangle$ is adjusted to couple preferentially the ground state to $\vert \psi_{\xi=N}^{(1)}\rangle$.  This can be realized experimentally with state of the art SLM technologies.   In Fig.~\ref{fig:steps}(a) we study the populations in the first three manifolds   when evolved with $\mathcal{H}=\mathcal{H}_{\textrm{eff}}+\mathcal{H}_F$ for $\Omega t<1$.  We see  in the insert  that the population with $n_{\textrm{exc}}$ excitations denoted  $\textrm{Pop}^{(n_{\textrm{exc}})}$ varies like $(\Omega t)^{2n_{\textrm{exc}}}$.
As a consequence, the single photon population dominates when $\Omega t\ll1$. This is  the so-called linear regime studied in~\cite{rubies2022photon,ballantine2021quantum} that neglects the inherent nonlinearity of atoms.
In this work, we illuminate the system for longer times and reach a regime where  both $\textrm{Pop}^{(1)}$ and $\textrm{Pop}^{(2)}$ contribute to the dynamics (while keeping $\textrm{Pop}^{(3)}\ll1$ for the sake of computation time).
In this nonlinear regime, our protocol benefits both from the increase of the total population in the chain, and from the possibility to harness the entanglement of doubly-excited states.
Moreover, the wavefront shaping of the incident beam proposed in this work simplifies the dynamics lowering the number of populated eigenmodes.  Indeed, we plot in Fig.~\ref{fig:steps}(b)  the projection of the density matrix onto each singly-excited eigenmode  and observe that $\textrm{Pop}^{(1)}$ is fully carried by the superradiant state $\vert \psi_{\xi=N}^{(1)}\rangle$ during the illumination: $\Gamma_0t<\Gamma_0t_1$. 
In the second manifold, we observed numerically that  the fermionic combination of the two most superradiant states  $\vert \psi^{(2)}_{\xi=(N-1,N)}\rangle$ largely dominates and contributes to $\sim 70\%$ of  $\textrm{Pop}^{(2)}$  (see appendix C for an analytical justification).

\begin{figure*}   
\begin{center}     
\includegraphics[width=0.9\linewidth]{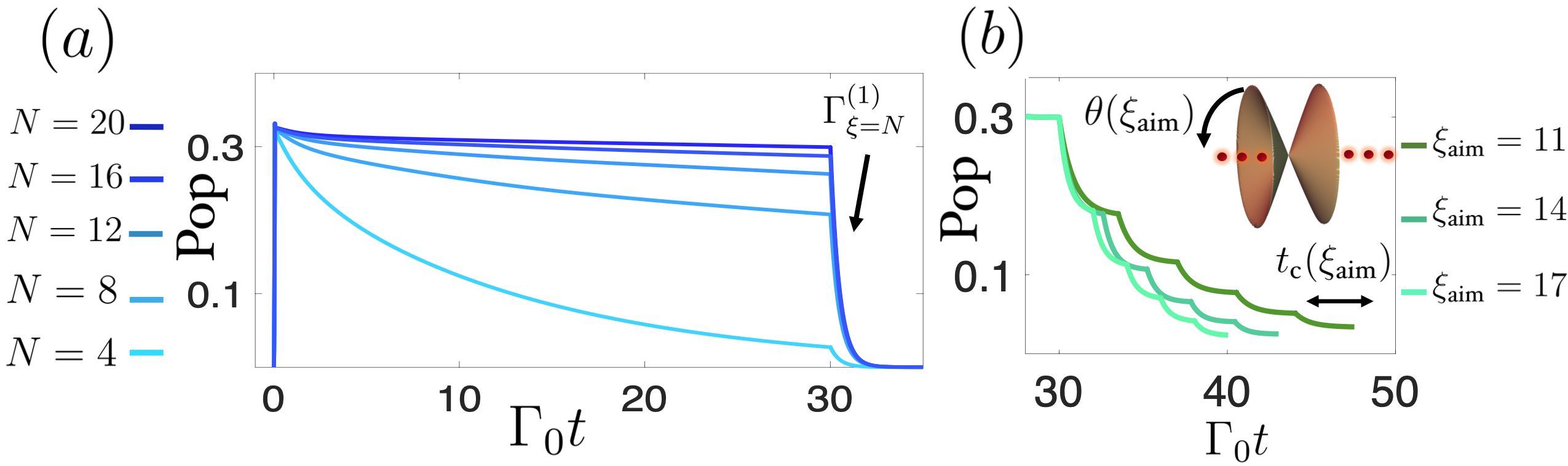} 
\end{center}
   \caption{a) Population in the chain as a function of time during the  protocol for different $N$ (indicated by different colors) and with $\Omega t_1=0.6$. (b) Emission step of the protocol with a control of the emitted mode $\xi_{\textrm{aim}}$. The emission process is split in $N_{\textrm{c}}\sim5$ cycles of duration $t_c(\xi_{\textrm{aim}})\sim 2/\Gamma_{\xi_{\textrm{aim}}}^{(1)}$. As represented in insert, each mode emits in a cone  of angle $\theta(\xi_{\textrm{aim}})$ with respect to the atomic chain. The parameters are: $a/\lambda_{eg}=0.35$, $N=20$  and the polarization is linear and parallel to the chain. The M.E equation is numerically truncated to $n_{\textrm{exc}}\leq2$.}
 \label{fig:memory}   
\end{figure*}

\subsection{Transfer to the most subradiant modes.}
After an illumination time $t_1$, we stop the driving  field and turn on the PVD: $\mathcal{H}=\mathcal{H}_{\textrm{eff}}+\mathcal{H}_S$.  
Since the  eigenmodes of $\mathcal{H}_{\textrm{eff}}$ are different from the eigenmodes of the total hamilatonian $\mathcal{H}$, the application of the PVD induces a transfer of population between the different $\vert \psi_{\xi}^{(1)}\rangle$.
In the specific case of  the staggered pattern: $\Delta_n=(-1)^n\Delta$, the  matrix representation of $\mathcal{H}_S$ in the singly-excited state basis of $\mathcal{H}_{\textrm{eff}}$  is perfectly anti-diagonal [see Fig.~\ref{fig:PVD}(a)].
This means that the application of this PVD induces a one-to-one coupling between $\vert \psi_{\xi}^{(1)}\rangle$ and $\vert \psi_{N+1-\xi}^{(1)}\rangle$~\cite{rubies2022photon}. 
More precisely, in the reduced basis built with $\{\vert \psi_{\xi}^{(1)}\rangle, \vert \psi_{N+1-\xi}^{(1)}\rangle \}$ the coupling matrix writes:
\begin{equation}
\mathcal{H}_{S}=
\begin{pmatrix}
0 & \Delta\\
\Delta& 0 
\end{pmatrix}.
\end{equation}

In Fig.~\ref{fig:steps}(b), we plot  the projection of the density matrix onto each singly-excited eigenmode as a function of time
during the first two steps of the protocol.
At the beginning of the transfer ($t=t_1$), $\textrm{Pop}^{(1)}$ is  carried by $\vert \psi_{\xi=N}^{(1)}\rangle$.
 From $t_1$ to $t_2$, we apply  $\mathcal{H}_{\textrm{S}}$ and observe a perfect transfer of the population to the most subradiant eigenmode $\vert \psi_{\xi=1}^{(1)}\rangle$.
 In the limit of a strong detuning:  $\Delta \gg \vert \omega^{(1)}_{\xi=N}\vert $ we can show that this transfer can be realized in a time $t_{t}=t_2-t_1=\pi/(2\Delta)\ll 1/\Gamma_{\xi=N}^{(1)}$ small enough to neglect radiative loss.

Let us now  push the analysis of the transfer to the two-photon population in the nonlinear regime.  In this case, the coupling matrix in the complete doubly-excited state basis is harder to interpret visually. However, for a given initial state  labeled with $\xi^{(2)}=(\xi_1,\xi_2)$, the evolution happens in a space of dimension 4 expressed in terms of the labels $\xi^{(2)}$ for simplicity:  $\{(\xi_1,\xi_2),(N+1-\xi_1,\xi_2),(\xi_1,N+1-\xi_2),(N+1-\xi_1,N+1-\xi_2)\}$.  
In this reduced basis, the transfer matrix has the form:
\begin{equation}
\mathcal{H}_{S}=·
\begin{pmatrix}
0 & -\Delta& \Delta& 0\\
-\Delta & 0 &0& \Delta\\
\Delta& 0 &0& -\Delta\\
0& \Delta & -\Delta &0
\end{pmatrix}.
\end{equation}
Since $70 \%$ of the doubly-excited population  is  contained into the most superradiant doubly-excited state $\xi^{(2)}=(N-1,N)$ at the beginning of the transfer,  we can restrict the discussion to the evolution of this  specific eigenmode. The application of the staggered pattern $\mathcal{H}_S$ first couples equally the most superradiant doubly-excited state  to $\xi^{(2)}=(1,N-1)$  and $\xi^{(2)}=(2,N)$. We note that these states are built with the anti-symmetric product of  one super and one subradiant singly-excited eigenstates and could be used for the heralded creation of subradiant states. Those states are in turn coupled to $\xi^{(2)}=(1,2)$ which is the most subradiant doubly-excited state. In order to illustrate this discussion, we plot in  Fig.~\ref{fig:steps}(c) the projection of the density matrix onto those four doubly-excited states.   Note that there is a factor of two between the projection of the density matrix onto a doubly-excited state and the atomic population that it induces as we deal with doubly-excited states.
At the beginning of the transfer, we see in  Fig.~\ref{fig:steps}(c) that the projection of the density matrix is maximal onto the most superradiant doubly-excited state $\vert \psi^{(2)}_{\xi=(N-1,N)}\rangle$. It is then efficiently transferred to  $\vert \psi^{(2)}_{\xi=(1,2)}\rangle$  in  the exact same time $t_{t}=\pi/(2\Delta)$ than derived for singly-excited states.

The application of the staggered pattern thus permits the coherent manipulation of doubly-excited states by switching their singly-excited components from superradiant to subradiant forms and vice-versa.  We believe that the same phenomenon happens for larger excitation number ($n_{\textrm{exc}}>2$) as soon as the generalization of the fermionic ansatz is valid. This work thus enlarges the  coherent control of collective states using the staggerred pattern~\cite{rubies2022photon,ballantine2021quantum} to higher manifolds.

\begin{figure*}   
\begin{center}     
\includegraphics[width=1\linewidth]{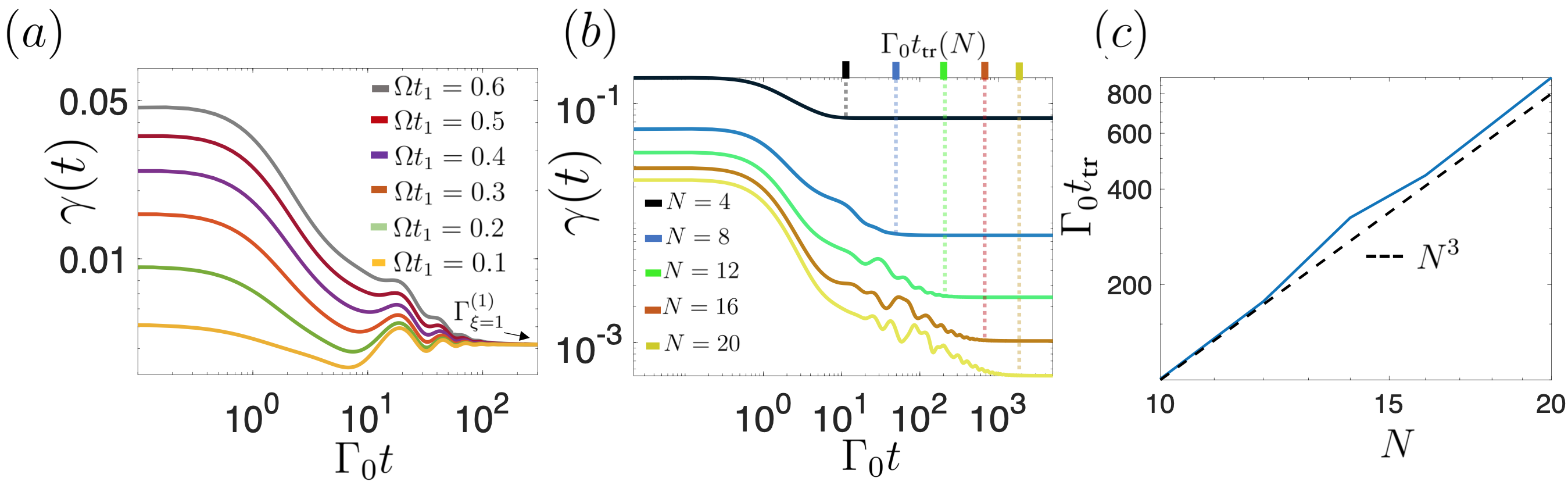} 
\end{center}
   \caption{ a) Numerical calculation of $\gamma(t)$ during the storage for various $\Omega t_1$ for a fixed $N=10$.  We highlight with a black arrow the long time value of $\gamma(t)=\Gamma_{\xi=1}^{(1)}$.  In (b), we fix $\Omega t_1=0.6$ and vary $N$. We project on top of the figure  the value of the transition time between the "short" and the "long" time behavior: $\Gamma_0 t_{\textrm{tr}}(N)$.  In (c)  we plot  $\Gamma_0 t_{\textrm{tr}}$ extracted from the simulations as a function of $N$ and compare it with a $N^3$ scaling.  The  parameters are: $a/\lambda_{eg}=0.35$, $n_{\textrm{exc}}\leq 2$, $N=10$ [in (a)] and $\Omega t_1=0.6$ in (b,c). The atomic  polarization is linear and parallel to the chain. In (a,b) the origin of time is taken at the beginning of the storage.}
 \label{fig:tsca}   
\end{figure*}

\subsection{Storage.}

After a transfer time $t_{t}=\pi/(2\Delta)$, we turn off $\mathcal{H}_S$  and let the system evolve freely for sufficiently long times such that the loss term in Eq.~(\ref{eq:Mastereq}) plays a role in the dynamics.
In Fig.~\ref{fig:memory}(a), we plot the total population in the chain during the complete protocol for various  $N$ indicated by different colors. 
Due to the open character of the system, we observe  a decrease of the atomic population during the storage ($t_2< t<30\Gamma_0$), which reveals that photons are emitted by the atomic chain. Importantly,  we observe that the storage of the atomic population increases with the system size $N$.

To discuss this first observation, we simplify the problem  neglecting the  $\sim 30\%$ of $\textrm{Pop}^{(2)}$ not contained into $\vert \psi^{(2)}_{\xi=(1,2)}\rangle$ and approximate the state of the system at the end of the transfer by the pure state $\vert \psi (t_2)\rangle\propto (\Omega t_1)\vert \psi^{(1)}_{\xi=1}\rangle+ \frac{(\Omega t_1)^2}{\sqrt{2}}\vert \psi^{(2)}_{\xi=(1,2)}\rangle$.
Then, we follow~\cite{henriet_clock} and use a rate model (R.M) described in  appendices B and D that simplifies the analysis.
Doing so, we restrict the the dynamics under free evolution in the presence of loss to a subspace built by the \emph{last two survivors} $\vert \psi^{(1)}_{\xi=1}\rangle$ and  $\vert \psi^{(2)}_{\xi=(1,2)}\rangle$. This simplification leads to the following expression for the total atomic population:
\begin{equation}
\textrm{Pop}(t)=\left(\Omega t_{1}\right)^2e^{\left(-\Gamma_{\xi=1}^{(1)}t\right)}+\left(\Omega t_{1}\right)^4e^{\left(-\Gamma_{\xi=(1,2)}^{(2)}t\right)}
\label{eq:Popana}
\end{equation} 
where the first term of the right hand side represents $\textrm{Pop}^{(1)}$ and the second term $\textrm{Pop}^{(2)}$.
 Since $\vert \psi^{(1)}_{\xi=(1)}\rangle$ and $\vert \psi^{(2)}_{\xi=(1,2)}\rangle$ are subradiant, both $\Gamma_{\xi=1}^{(1)}$ and $\Gamma_{\xi=(1,2)}^{(2)}$  decrease as $1/N^3$. Injecting those scalings into Eq.~(\ref{eq:Popana}) directly permits to recover the improvement of the storage of the total population when the system size increases  observed numerically in Fig.~\ref{fig:memory}(a).

Next, we push the analysis further comparing the rate of emission of photons coming from singly and doubly-excited states.
To do so we study  the instantaneous decay rate~\cite{rubies2023dynamic}:
\begin{equation}
    \gamma(t)=-\textrm{Pop}'/\textrm{Pop}
    \label{eq:localder}
\end{equation}
during the storage,  for various $N$ and different values of $\Omega t_1$.  
In Fig.~\ref{fig:tsca}(a) we plot the numerical value of $\gamma(t)$  for $N=10$ and $\Omega t_1$ varying from $0.1$ to $0.6$. We see that whatever the value of $\Omega t_1$, $\gamma(t)$ converges at "long" times towards a plateau whose value is given by $\Gamma_{\xi=1}^{(1)}$.  
This means that the emission of photons at "long" times is dominated by single photons radiated by the most subradiant singly-excited state. 

Next we turn to the study of the "short" time regime of $\gamma(t)$ observed in Fig.~\ref{fig:tsca}(a) where $\gamma(t)$ varies with $\Omega t_1$.
To understand this point, we inject Eq.~\ref{eq:Popana} into Eq.~\ref{eq:localder} and obtain a simplified expression of $  \gamma(t)$ predicted by the R.M. It writes:
\begin{equation}
    \gamma(t)=\Gamma_{\xi=1}^{(1)}\frac{e^{\left(-\Gamma_{\xi=1}^{(1)}t\right)}+\kappa(\Omega t_1)^2e^{\left(-\Gamma_{\xi=(1,2)}^{(2)}t\right)}}{e^{\left(-\Gamma_{\xi=1}^{(1)}t\right)}+(\Omega t_1)^2e^{\left(-\Gamma_{\xi=(1,2)}^{(2)}t\right)}},
    \label{eq:localdermodel}
\end{equation}
with $\kappa>1$ the proportionality  factor between  $\Gamma_{\xi=1}^{(1)}=\alpha_1 N^{-3}$ and $\Gamma_{\xi=(1,2)}^{(2)}=\kappa \alpha_1 N^{-3}$ which comes from the fermionic ansatz~\cite{asenjo2017exponential}.
In Fig.~\ref{fig:tfit}(b), we compare Eq.~(\ref{eq:localdermodel}) with the exact calculation of $\gamma(t)$ using the M.E and observe that Eq.~(\ref{eq:localdermodel})  slightly underestimates the instantaneous decay at short times. This comes  from the neglect of the $\sim 30\%$ of $\textrm{Pop}^{(2)}$ not contained into $\vert \psi^{(2)}_{\xi=(1,2)}\rangle$.  Nonetheless the R.M still captures the increase of $\gamma(0)$ when $\Omega t_1$ increases, and shows that this effect comes from the competition between the singly and doubly-excited decay. 
More importantly, we observe in  Fig.~\ref{fig:tfit}(b) that  Eq.~(\ref{eq:localdermodel}) properly captures the dynamical transition  between the short and long time  regimes of $\gamma(t)$. Hence the R.M permits to set the boundary between a short time regime where the emission is dominated by   $\textrm{Pop}^{(2)}$ and a long time regime where it is dominated by $\textrm{Pop}^{(1)}$.
 To do so, we introduce the transition time $t_{\textrm{tr}}$ such as  the time needed for $\gamma(t)$ to converge towards  $\Gamma_{\xi=1}^{(1)}$. We obtain its analytical expression using Eq.~\ref{eq:localdermodel} using the condition that  the weight of the single and  two-photon decay should be equal at this specific time. We obtain:
\begin{equation}
t_{\textrm{tr}}=\frac{\textrm{log}(\kappa (\Omega t_{1})^2)}{(\kappa-1)\alpha_1}N^3
\label{eq:ttr}
\end{equation} 
with $\kappa=\Gamma_{\xi=(1,2)}^{(2)}/\Gamma_{\xi=(1)}^{(1)}$   the ratio between the two decay rates almost constant to a value around $9$ for large systems (see Appendix D).

In Fig.~\ref{fig:tsca}(b), we fix $\Omega t_1=0.6$,  vary  $N$ from $4$ to $20$ and extract the numerical value of $t_{\textrm{tr}}$ that we project on top of the figure for the sake of clarity.
In Fig.~\ref{fig:tsca}(c), we plot the extracted value of $t_{\textrm{tr}}$ as a function of the system size and  confirm numerically the variation of   $t_{\textrm{tr}}\sim N^3$  predicted by Eq.~\ref{eq:ttr}. 
This scaling shows that doubly-excited states dominate the emission for a time that strongly increases with the system size.
Besides, Eq.~\ref{eq:ttr}  tells us that $t_{\textrm{tr}}$ only exists if $\Omega t_1>\kappa^{-1/2}$ (due to the $\textrm{log}$).
The condition  $\Omega t_1>\kappa^{-1/2}$   depicted in  Fig.~\ref{fig:steps}(a) thus sets the boundary between the linear and the nonlinear regime (in terms of illumination strength) to $\Omega t_1=1/\sqrt{\kappa}\sim 0.3$. 
This order of magnitude  is  confirmed numerically in  Fig.~\ref{fig:tsca}(a) in which we can estimate  the minimum value of $\Omega t_1$ such that  $\textrm{Pop}^{(2)}$ dominates the decay at short times to  $\sim 0.2$. 

To conclude this section, we proposed a simplified model of the dynamics during the storage  in order to compute  the minimal value of the illumination strength $\Omega t_1$ needed to access the onset of the non-linear regime. In this regime, we showed that  the two-photon population  dominates the emission of photons for a duration that strongly increases with the system size.

\subsection{Emission.}
After a sufficiently long storage time  such that  the two-photon population is completely negligible, the  remaining  population  is fully carried by the most subradiant singly-excited state  $\vert \psi^{(1)}_{\xi=1}\rangle$.
On demand,  one can turn on the PVD,  transfer back the   population to the most superradiant singly-excited mode in a time $t_{t}=\pi/(2\Delta)$, and then let evolve the state with $\mathcal{H}_{\textrm{eff}}$. Doing so, the  atomic population decays  at a rate $\Gamma_{\xi=N}^{(1)}$, as observed in Fig.~\ref{fig:memory}(a).
In order to control the emission rate: $\Gamma_{\xi_{\textrm{aim}}}^{(1)}$, the frequency $\omega_{\xi_{\textrm{aim}}}^{(1)}$ and the radiation pattern of the emitted single-photon, we introduce  a new sinusoidal PVD:
\begin{equation}
 \Delta_n(t)=\Delta\textrm{sin}\left(\frac{n\xi_{aim}\pi}{N+1}\right)
 \end{equation}
  that  couples the most subradiant eigenmode $\vert \psi_{\xi=1}^{(1)}\rangle$ to $\vert \psi^{(1)}_{\xi_{\textrm{aim}}}\rangle$.
As observed in Fig.~\ref{fig:PVD}(c) this coupling is not perfectly one-to-one, and rather looks like a $\textrm{sinc}$ function. In order to avoid second order coupling to subradiant states which would prevent the efficient emission of the photon, we   split the emission procedure in $N_{\textrm{c}}\sim 5$ cycles. 
Each cycle contains one  transfer step using the sinusoidal PVD of duration $t=\pi/(3\Delta)$ short enough for enabling first order coupling only, and one free evolution step of duration $t=2/\Gamma_{\xi_{\textrm{aim}}}^{(1)}$.
This method enables the efficient emission of the photon mostly through the mode indexed  $\xi_{\textrm{aim}}$ as shown in Fig.~\ref{fig:memory}(b).
We  compute the radiation pattern of the atomic ensemble using: $
\mathbf{E}(\mathbf{r})=\mu_0\omega_0^2\sum_{n=1}^N\overline{\mathbf{G}}(\mathbf{r},\mathbf{r}_n,\omega_0)\mathbf{p}_n\langle \sigma_{ge}^n\rangle
$ with $\langle \sigma_{ge}^n\rangle=\textrm{Tr}\left(\sigma_{ge}^n\rho\right)$ obtained from the resolution of eq.~(\ref{eq:Mastereq}). 
The radiation pattern of one typical superradiant mode indexed with $\xi_{\textrm{aim}} \neq N$ is given the inset of Fig.~\ref{fig:memory}(b). For atoms linearly polarized along the chain, we see that the mode  with label $\xi_{\textrm{aim}}$ emits in a cone of angle $\theta(\xi_{\textrm{aim}})$ that directly depends on $k_z$: $\theta(\xi_{\textrm{aim}})=\textrm{atan}(k_\perp/k_z)$, with $k_z=\frac{\pi(N+1-\xi_{\textrm{aim}})}{a(N+1)}$ and $k_\perp=\sqrt{\omega_{eg}^2/c^2-k_z^2}$.
Perfect retrieval of the emitted single-photon can thus be realized with two lenses of proper NA.

\section{Conclusion}

In summary, we studied theoretically  a  protocol that enables on-demand absorption, storage, and re-emission of an incident field
 using subradiant and superradiant states of an atomic chain.  The three-level nature of the atoms  allows both their trapping and manipulation using light shifts with a resolution  smaller than $\lambda_{eg}$.  This permits the determistic coupling to subradiant states, which are important resources for metrology~\cite{ostermann2013protected,facchinetti2018interaction} or quantum computing~\cite{ wild2018quantum,guimond2019subradiant}.
Valid beyond the linear regime, our theoretical  analysis  provides an intuitive picture of the dynamics in terms of  singly and doubly-excited states.
For the first time to our knowledge, we discuss the manipulation of multiply excited states using tailored PVDs combined with shaped incident fields. 
We envision in the future  to study the efficient excitation of highly excited states in order to control  the quantum correlations of light emitted by the array~\cite{masson2020many,richter2023collective}.  
 Quantum metamaterials~\cite{bekenstein2020quantum,solntsev2021metasurfaces}, such as $2$D arrays  have already been shown to act as a  subradiant mirrors~\cite{bettles2016enhanced,rui2020subradiant,srakaew2023subwavelength} or proposed 
to construct lossless $1$D
atomic waveguides~\cite{asenjo2017exponential,masson2020atomic}. Adjusting  their many-body functionalities on demand, in a time shorter than the emission time 
is a largely unexplored avenue that would enlarge their potential as versatile light-matter interfaces.

\section*{Acknowledgements}
We acknowledge Ilan Shlesinger and Jean-Paul Hugonin for important discussions.
A.B. and I.F-B acknowledge funding from the European Research Council (Grants. No. 101018511, ATARAXIA, 101039361, CORSAIR)
and by the Agence National de la Recherche (ANR, project DEAR).

\begin{figure}   
\begin{center}     
\includegraphics[width=0.8\linewidth]{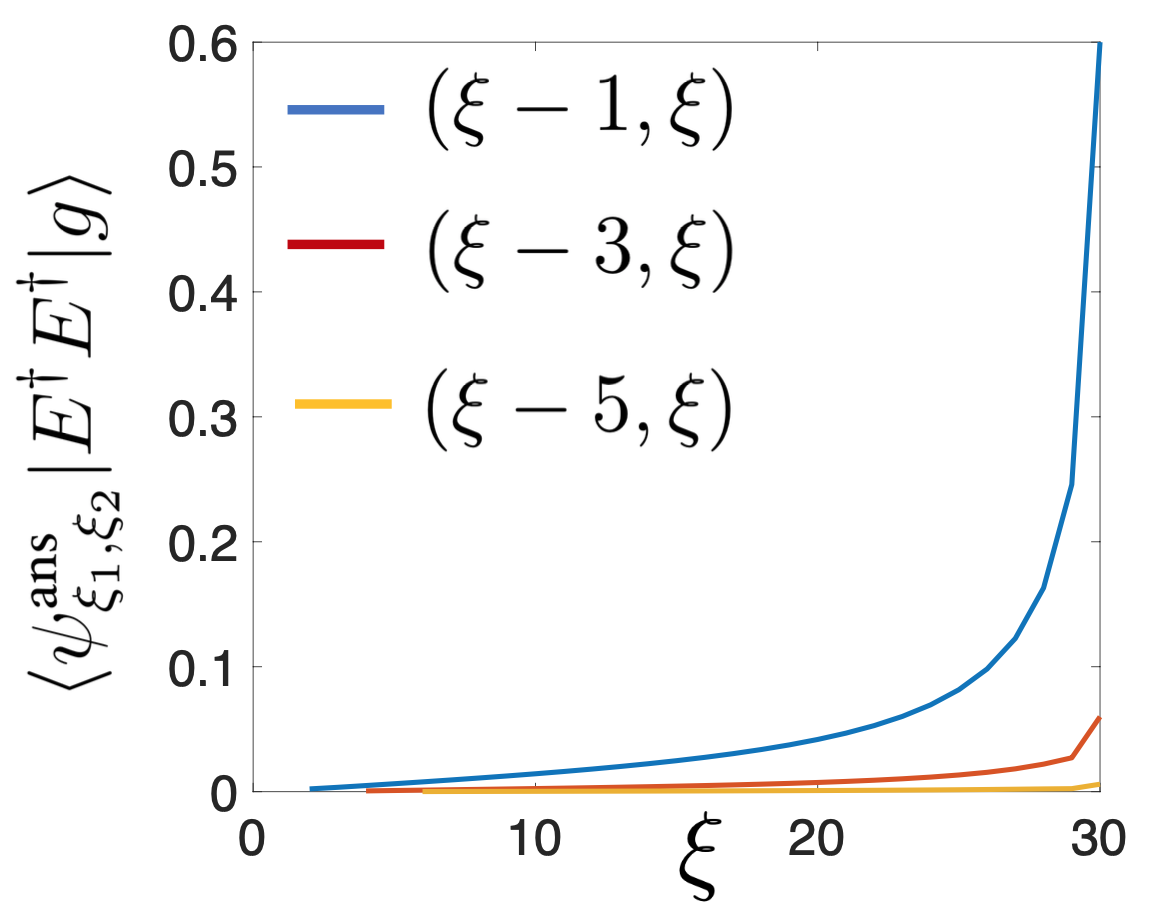} 
\end{center}
   \caption{ Numerical calculation of $\langle \psi^{\textrm{ans}}_{\xi_1,\xi_2} \vert E^\dagger E^\dagger\vert g \rangle/(\mathcal{N}\Omega^2)$ given  Eq.~(\ref{eq:coupling}). We study three different types of doubly-excited states built with the anti-symmetric product of singly-excited eigenmodes with labels $(\xi-1,\xi)$ (blue),  $(\xi-3,\xi)$ (red) and  $(\xi-5,\xi)$  (yellow) as a function of $\xi$. The system size is $N=30$, $\Omega=1$, and the eigenmodes have been computed with an atomic polarization parallel to the chain. }
 \label{fig:fieldcoupling}   
\end{figure}

\appendix\section{\\Collective properties of singly-excited states}

\begin{figure*}   
\begin{center}     
\includegraphics[width=1\linewidth]{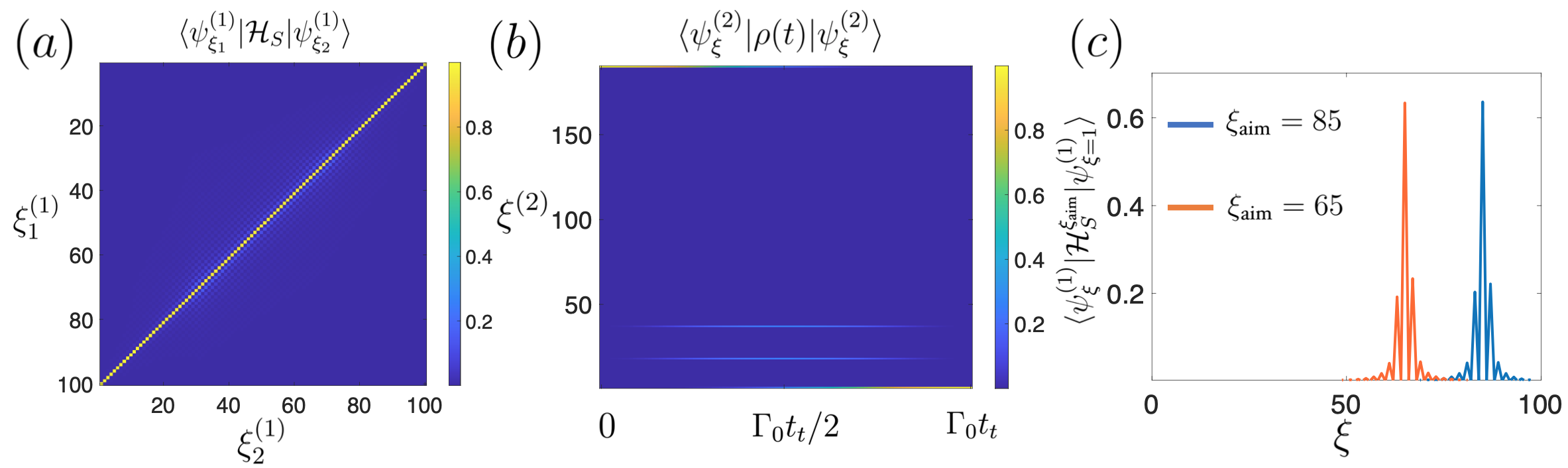} 
\end{center}
   \caption{a) Matrix element of $\mathcal{H}_{\textrm{S}}$ (with a staggered pattern) in the basis built with the different eigenmodes $\vert \psi_\xi^{(1)}\rangle$. b) Projection of the time evolved density matrix onto the doubly-excited eigenmodes during the transfer with the most superradiant doubly-excited state.  We observe that the two-photon population starts in $\xi^{(2)}=\binom{N}{2}=190$ before being split into two intermediate states $\xi^{(2)}=N=20$ and $\xi^{(2)}=N+N-1=39$ (half super and half subradiant) and terminates the transfer into $\xi^{(2)}=1$. In this plot, $\xi^{(2)}$ is not a 2D vector but a number that sorts the doubly-excited eigenmodes with respect to their decay rate. c) Coupling parameter between $\vert \psi_{\xi=1}^{(1)}\rangle$ and $\vert \psi_{\xi}^{(1)}\rangle$ induced by the sinusoidal PVD: $\mathcal{H}_{\textrm{S}}^{\xi_\textrm{aim}}$ for  $\xi_\textrm{aim}=65$ (red) and $\xi_\textrm{aim}=85$ (blue). The parameter are $N=100$, $a=0.35\lambda_{eg}$ and $\Delta=\Gamma_0$ in (a,c). In (b) we considered $N=20$, $a=0.35\lambda_{eg}$ and $\Delta=100\Gamma_0$. The atomic polarization is parallel to the chain.}
 \label{fig:PVD}   
\end{figure*}

In this appendix, we remind the analytical expressions of $\Gamma^{(1)}_{k_z}$ and $\omega^{(1)}_{k_z}$ derived in the literature~\cite{asenjo2017exponential,shahmoon2017cooperative} in order to understand the collective properties of the eigenmodes of the effective Hamiltonian.
$\Gamma^{(1)}_{k_z}$ can be expressed in terms of the reciprocal lattice vectors $g_z$, and writes (for a linear atomic polarization along the chain):
\begin{equation}
\frac{\Gamma^{(1)}_{k_z}}{\Gamma_0}=\frac{3\pi}{2k_0a}\sum_{\vert k_z+g_z\vert <k_0}\left(1-\frac{(k_z+g_z)^2}{k_0^2}\right).
\label{eq:anagamma}
\end{equation}
The analytical expression of $\omega^{(1)}_{k_z}$ can be obtained using the mathematical function $\operatorname{Li}_s(z)=\sum_{l=1}^\infty z^le^{-s}$, $Z_1=e^{i(k_0+k_z)a}$ and  $Z_2=e^{i(k_0-k_z)a}$. It writes:
\begin{equation}
\begin{split}
\frac{\omega^{(1)}_{k_z}}{\Gamma_0}=&-\frac{3\pi}{2k_0^3a^3}\left(\operatorname{Li}_3(Z_1)+\operatorname{Li}_3(Z_2)\right.\\
&\left.-ik_0a[\operatorname{Li}_2(Z_1)+\operatorname{Li}_2(Z_2)]\right).
\end{split}
\label{eq:anaomega}
\end{equation}
We stress here that those quantities depends  on the atomic polarization. The numerical results  presented in this work have been derived for a linear polarization along the chain. However, they can be easily extended to the same system with  circular atomic polarization  simply by adapting the exact value of the decay rates and the frequency shifts~\cite{asenjo2017exponential,shahmoon2017cooperative}. For  an atomic polarization linear and orthogonal  to the chain (along $x$ for instance), the link between $\xi$ and $k_z$ is more involved, but the physics of super and subradiance remains the same.  Namely, spin waves with a wave vector along the chain larger than $\omega_{eg}/c$ cannot couple to electromagnetic waves  in vacuum and are strongly subradiant.  On the opposite, superradiant modes are associated with small values of $k_z<\omega_{eg}/c$.

\section{Decay of the doubly-excited eigenmodes}

In this appendix, we remind some properties of the most super and subradiant doubly-excited states  useful to understand the population  dynamics during the protocol in the nonlinear regime.
The most subradiant (respectively superradiant) doubly-excited eigenmode is built as a fermionic product of singly-excited states with indexes  $(\xi_1=1,\xi_2=2)$ [respectively $(\xi_1=N-1,\xi_2=N)$]. As a direct consequence, the most superradiant doubly-excited eigenmode mostly decays into the singly-excited states of indexes  $\xi_1=N-1$ and $\xi_2=N$ while the most subradiant decays into $\xi_1=1$ and $\xi_2=2$ \cite{henriet_clock}.
The reduction of the number of states involved in the decay process enables the use of a  simplified  rate model  involving only 3 different states: $\{\vert \psi_{\xi_1}^{(1)} \rangle,\vert \psi_{\xi_2}^{(1)} \rangle,\vert \psi_{\xi}^{(2)}\rangle \}$:
\begin{equation}
    \begin{cases}
      \mathrm{d} c^{(2)}_{\xi}/ \mathrm{d}t= -\Gamma_{\xi}^{(2)} c^{(2)}_{\xi}\\
     \mathrm{d} c^{(1)}_{\xi_1}/ \mathrm{d}t= -\Gamma_{\xi_1}^{(1)} c^{(1)}_{\xi_1}+\gamma_{\xi_1,\xi}c^{(2)}_{\xi}\\
    \mathrm{d} c^{(1)}_{\xi_2}/ \mathrm{d}t= -\Gamma_{\xi_2}^{(1)} c^{(1)}_{\xi_2}+\gamma_{\xi_2,\xi}c^{(2)}_{\xi}
    \end{cases}\
    \label{eq:RM}
\end{equation}
with $c^{(2)}_{\xi}$ the population of the doubly-excited mode,  $c^{(1)}_{\xi_1}$ and $c^{(1)}_{\xi_2}$ the population of the singly-excited modes into which it decays.
 The decay rate from  state $\vert \psi_{\xi'}^{(2)}\rangle$ to state $\vert \psi_\xi^{(1)}\rangle$ can be computed using $\gamma_{\xi,\xi'}=\textrm{Tr}\left(\vert \psi_\xi^{(1)}\rangle\langle \psi_\xi^{(1)}\vert \mathcal{J}\vert \psi_{\xi'}^{(2)}\rangle\langle \psi_{\xi'}^{(2)}\vert \right)$, with $\mathcal{J}=\sum_{m,n}\Gamma_{m,n}\sigma_{eg}^m\rho\sigma_{ge}^n$  the recycling operator of the M.E given in Eq. (1) of the main text.

\section{Details of the steps of the protocol}
\subsection{Coupling with the field}

\begin{figure*}   
\begin{center}     
\includegraphics[width=1\linewidth]{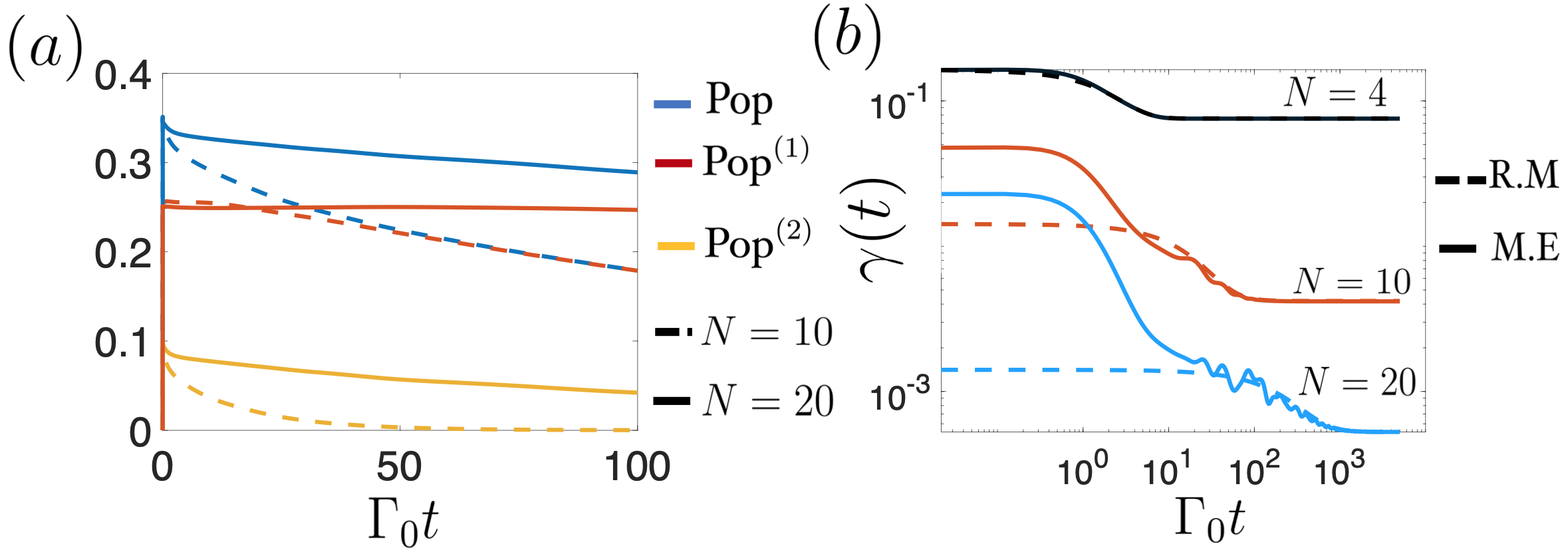} 
\end{center}
   \caption{a) Total population (blue), population in the fist (red) and second (yellow) manifolds  as a function of time for $N=10$ (dashed) and $N=20$ (solid lines)
   b)  Numerical calculation $\gamma(t)$ during the storage for various $N$ and $\Omega t_1=0.6$. In solid lines we plot the results given by the M.E and in dashed the results given by the R.M. The Hilbert space has been truncated to $n_{\textrm{exc}}\leq 2$, we considered atomic  polarization  linear and parallel to  the chain, and an interatomic distance  $a/\lambda_{eg}=0.35$. In (b) the origin of time is taken at the beginning of the storage ($t=t_2$).} 
 \label{fig:tfit}   
\end{figure*}

We chose the field operator $E^\dagger=\Omega\sum_ic_{\xi=N}^{(1)}(i)\sigma_{eg}^i$  in order to couple the ground state to $\vert \psi_{\xi=N}^{(1)}\rangle$ efficiently. Indeed, one can  check numerically or analytically that $\langle\psi^{(1)}_{\xi} \vert E^\dagger\vert g\rangle=\Omega\delta\left(\xi-N\right)$. However we do not know \emph{a priori} which doubly-excited eigenstates are populated when the field operator is applied twice to the ground state and creates $E^\dagger E^\dagger\vert g\rangle=\Omega^2\sum_{i<j}c_{\xi=N}^{(1)}(i)c_{\xi=N}^{(1)}(j)\sigma_{eg}^i\sigma_{eg}^j\vert g\rangle$.
To identify this state, we take its dot product with every two-photon ansatz states:  $\langle \psi^{\textrm{ans}}_{(\xi_1,\xi_2)} \vert E^\dagger E^\dagger\vert g \rangle $ using:
\begin{equation}
\begin{split}
\vert \psi _{(\xi_1,\xi_2)} ^{\textrm{ans}}\rangle= &\mathcal{N}\sum_{m<n}\left[ c_{\xi_1}^{(1)}(m)c_{\xi_2}^{(1)}(n)-c_{\xi_2}^{(1)}(m)c_{\xi_1}^{(1)}(n)\right]\\
&\times \sigma_{eg}^m\sigma_{eg}^n\vert g\rangle,
\end{split}
\end{equation}
with $\mathcal{N}$ a normalization factor and $c_{\xi}^{(1)}(i)=\sqrt{\frac{2}{N+1}}\textrm{sin}\left[\frac{(N+1-\xi)i\pi}{N+1}\right]$ the singly-excited eigenmode ansatz.
This leads to a rather long expression:

\begin{equation}
\begin{split}
&\langle \psi^{\textrm{ans}}_{\xi_1,\xi_2} \vert E^\dagger E^\dagger\vert g \rangle=\mathcal{N}\sum_{i<j}\frac{4\Omega^2}{(N+1)^2}\textrm{sin}\left[\frac{i\pi}{N+1}\right]\textrm{sin}\left[\frac{j\pi}{N+1}\right] \\
&\times\left[ \textrm{sin}\left[\frac{(N+1-\xi_1)i\pi}{N+1}\right]\textrm{sin}\left[\frac{(N+1-\xi_2)j\pi}{N+1}\right]  \right.\\
&\left. -\textrm{sin}\left[\frac{(N+1-\xi_2)i\pi}{N+1}\right]\textrm{sin}\left[\frac{(N+1-\xi_1)j\pi}{N+1}\right]  \right] 
\end{split}
\label{eq:coupling}
\end{equation}
that we evaluate numerically in Fig.~\ref{fig:fieldcoupling}. We  show that, when applied two times on the ground state, the field preferentially couples to  $\vert \psi^{\textrm{ans}}_{(\xi_1=N-1,\xi_2=N)}\rangle$: the most superradiant doubly-excited state. 
One should note that a part of the two-photon population is carried by other modes  ($\sim 30\%$ of the  two-photon population observed numerically). This $ 30\%$ of $\textrm{Pop}^{(2)}$ is responsible for the discrepancy observed at  short times in the evaluation of $\gamma(t)$ using the R.M and the M.E observed in Fig.~\ref{fig:tfit}(b).

\subsection{Modes transfer using PVD}
In this subsection, we provide additional details about mode transfer using PVD. In Fig.~\ref{fig:PVD}(a) we represent $\vert \langle \psi_{\xi_1}^{(1)}\vert \mathcal{H}_{\textrm{S}}\vert \psi_{\xi_2}^{(1)}\rangle\vert$ and observe an almost perfect anti-diagonal matrix. This means that, the evolution of an eigenmode with label $\xi$ of $\mathcal{H}_{\textrm{eff}}$ under the application of the staggered pattern happens in a space of dimension $2$: $\{\vert \psi_{\xi}^{(1)}, \vert \psi_{N+1-\xi}^{(1)}\rangle \}$.

In the second manifold, we choose not to  represent the matrix elements $\vert \langle \psi_{\xi_1}^{(2)}\vert \mathcal{H}_{\textrm{S}}\vert \psi_{\xi_2}^{(2)}\rangle\vert$ as they are difficult to interpret visually. Instead, we plot in Fig.~\ref{fig:PVD}(b) the projection of the density matrix onto each doubly-excited eigenmode during the transfer.  In this plot, $\xi^{(2)}$ is not a 2D vector but a number that sorts the doubly-excited eigenmodes with respect to their decay rate. The initial state is the most superradiant doubly-excited state associated with $\xi^{(2)}=\binom{N}{2}=190$ (for $N=20$ atoms). It is then equally transferred to the doubly-excited state of index $\xi^{(2)}=N=20$ (built with the product of $\vert \psi_{\xi=1}^{(1)} \rangle$ and $\vert \psi_{\xi=N}^{(1)} \rangle$) and the doubly-excited state of index $\xi^{(2)}=N+(N-1)=39$ (built with the product of $\vert \psi_{\xi=2}^{(1)} \rangle$ and $\vert \psi_{\xi=N-1}^{(1)} \rangle$). Eventually, those two modes are coupled to the most subradiant doubly-excited state associated with $\xi^{(2)}=1$ (built with the product of $\vert \psi_{\xi=1}^{(1)} \rangle$ and $\vert \psi_{\xi=2}^{(1)} \rangle$).

At the end of the storage, we want to transfer back the singly-excited subradiant state into the superradiant mode of our choice. Thus, we use the following sinusoidal PVD: $\mathcal{H}_{\textrm{S}}^{\xi_\textrm{aim}}=-\sum_{n=1}^N\Delta_n\sigma_{ee}^n$ with $\Delta_n=\Delta \textrm{sin}\left[n\xi_{\textrm{aim}}\pi/(N+1)\right]$. This detuning pattern mostly couples the most subradiant mode ($\xi=1$)  to the eigenstate with label $\xi_{\textrm{aim}}$ as observed in Fig.~\ref{fig:PVD}(c).
 However, since the coupling is not one to one (it rather looks like a sinc function), the emission protocol should be split into $N_{\textrm{c}}=5$ cycles. Each cycle contains one  transfer step using the sinusoidal PVD of duration $t=\pi/(3\Delta)$ and one free evolution step of duration $t=2/\Gamma_{\xi_{\textrm{aim}}}^{(1)}$.  With this method, we can funnel most of  the emission through the singly-excited  eigenmode  with label $\xi_{\textrm{aim}}$ and avoid residual coupling to subradiant eigenmodes.

\section{Last two survivors approximation}
In this section, we provide additional details about the analysis of  the role of  the two-photon population during the storage. As a first observation, we plot in Fig.~\ref{fig:tfit}(a) the total population (blue), the population of the single photon components (red) and two-photons components (yellow) for two different system sizes ($N=10$ in dashed and $N=20$ in solid lines) for $\Omega t_1=0.6$.   We observe that both the single and the two-photon components are better stored as $N$ increases and clearly observe that the two-photon population can not be neglected for "short"  times. 

 \begin{figure}   
\begin{center}     
\includegraphics[width=0.8\linewidth]{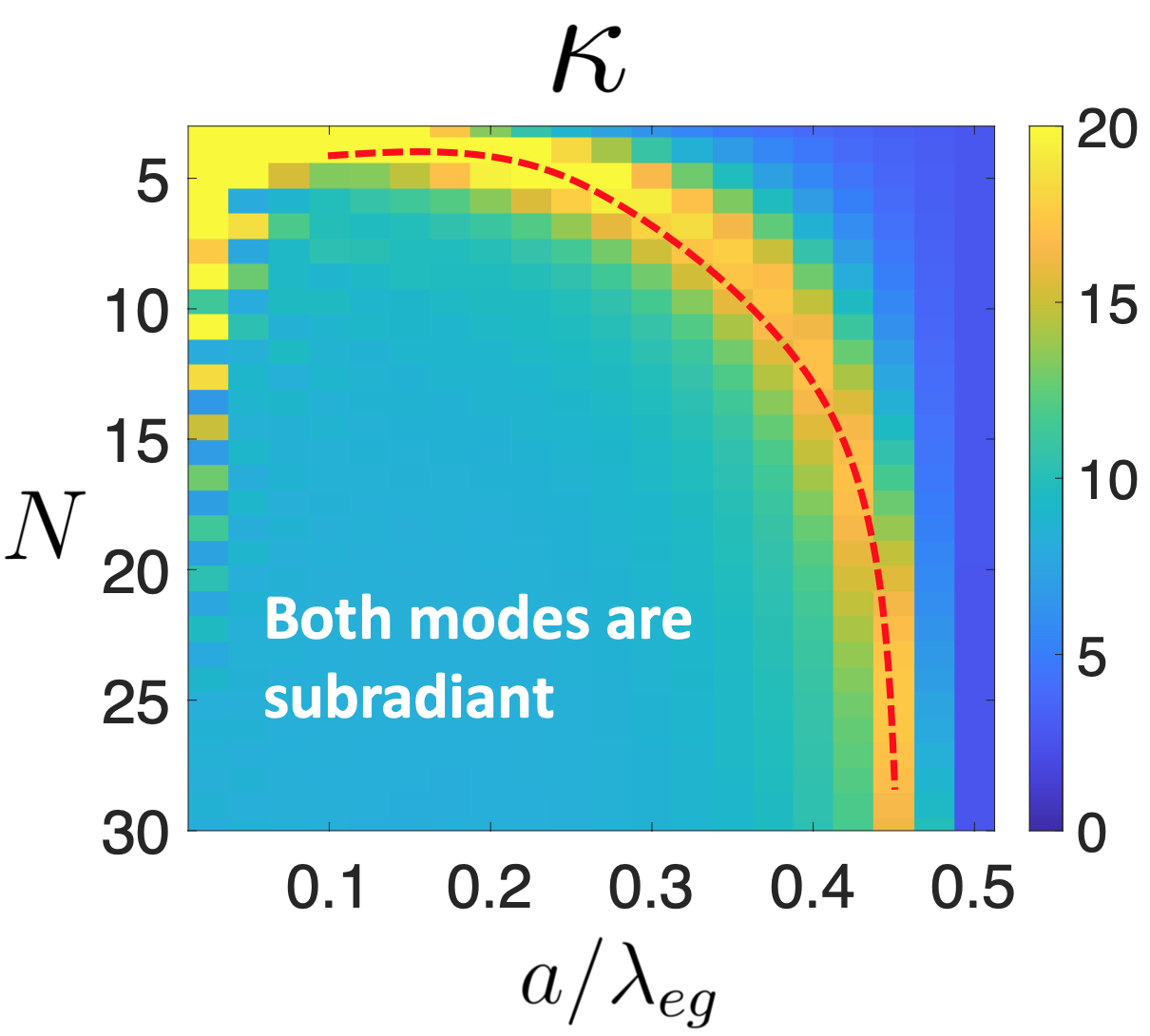} 
\end{center}
   \caption{a)  Numerical calculation of $\kappa=\Gamma_{\xi=(1,2)}^{(2)}/\Gamma_{\xi=1}^{(1)}$ as a function of $N$ and $a/\lambda_{eg}$.  In the regime where both modes are subradiant, $\kappa$ is found to be almost constant and close to $9$.  In dashed red, we highlight the parameters for which the most subradiant doubly-excited state  becomes superradiant (due to finite size effects) while the  singly-excited one remains subradiant. This results in an increase of $\kappa$. Outside  this domain, both modes are radiants. We considered atomic polarization linear and parallel to the chain.}
 \label{fig:kappa}   
\end{figure}

\begin{figure*}   
\begin{center}     
\includegraphics[width=1\linewidth]{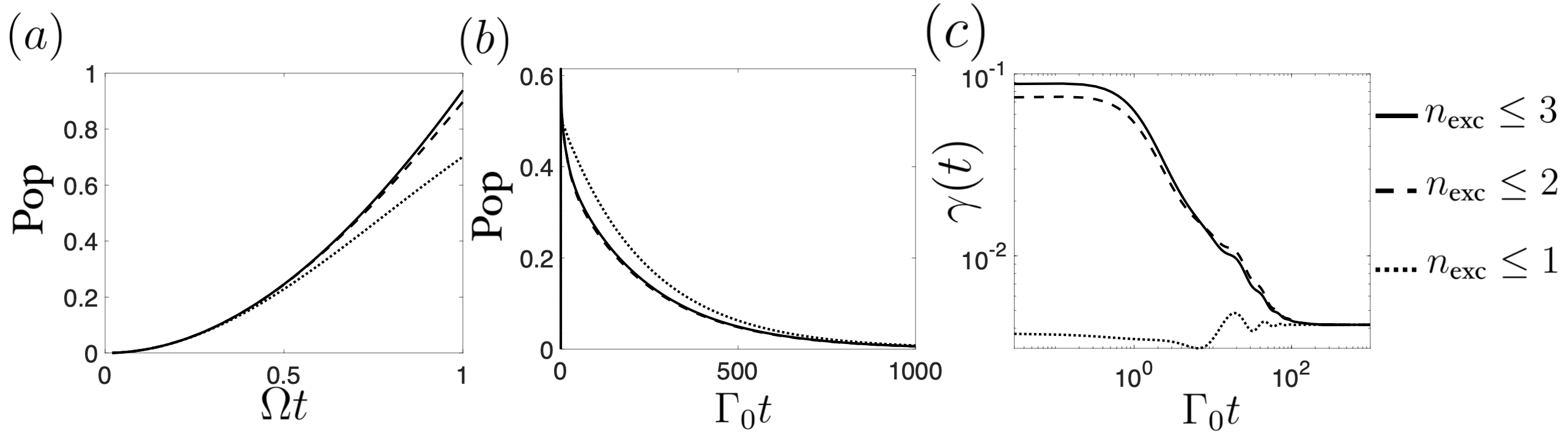} 
\end{center}
   \caption{a) Total atomic population in the chain for a numerical truncation of the M.E up to $n_{\textrm{exc}}=1$ (black dotted), $2$ (black dashed) and $3$ (solid black) during the illumination step in (a) and during the storage step of the protocol in (b).  The numerical parameters are $N=10$, $a/\lambda_{eg}=0.35$, $\Omega=100\Gamma_0$, $\Delta=100\Gamma_0$, $\Omega t_1=0.8$ and the polarization of the atoms is parallel to the chain. In (c) the origin of time is taken at the beginning of the storage ($t=t_2$).} 
 \label{fig:3phot}   
\end{figure*}

In Fig.~\ref{fig:tfit}(b) we compare the value of $\gamma(t)$ given by the R.M and by the M.E. We observe a good agreement for both the  long time value of $\gamma(t)$ and  $t_{\textrm{tr}}$ as discussed in the main text. This shows that we can use the reduced set of states of the R.M: $\{\vert \psi_{\xi=1}^{(1)}\rangle, \vert \psi_{\xi=2}^{(1)}\rangle,\vert \psi_{\xi=(1,2)}^{(2)}\rangle\}$ to compute the dynamics for  $t\geq t_{\textrm{tr}}$. To do so, we express the total population during the storage in the reduced basis:
\begin{equation}
\textrm{Pop}(t)=c_{\xi=1}^{(1)}(t)+c_{\xi=2}^{(1)}(t)+2c_{\xi=(1,2)}^{(2)}(t)
\label{eq:Pop2}
\end{equation}
with $c_{\xi=1}^{(1)}(t)$, $c_{\xi=2}^{(1)}(t)$ and $c_{\xi=(1,2)}^{(2)}(t)$ the population of the three modes involved. Their initial values at $t=0$ (beginning of the storage) depend on $\Omega t_1$: $c_{\xi=1}^{(1)}(0)=(\Omega t_1)^2$, $c_{\xi=2}^{(1)}(0)=0$ and $2c_{\xi=1}^{(2)}(0)\sim(\Omega t_1)^4$ and their dynamics can be solved numerically using eq.~\ref{eq:RM}.
In this specific study of the transition time $t_{\textrm{tr}}$, $c_{\xi=2}^{(1)}(t)$ plays a negligible role as  both its population and decay rate are low. We  thus reduce the analytical analysis to the \emph{last two survivors} $c_{\xi=1}^{(1)}(t)$ and  $c_{\xi=(1,2)}^{(2)}(t)$, neglect the filling of $c_{\xi=1}^{(1)}(t)$ due to the decay of $c_{\xi=(1,2)}^{(2)}(t)$ in order to obtain the simplified expression of the total population provided in Eq.~\ref{eq:Popana}.
Equation~\ref{eq:Popana} is written in terms of $\Gamma_{\xi=1}^{(1)}=\alpha_1 N^{-3}$ and $\Gamma_{\xi=(1,2)}^{(2)}=\kappa \alpha_1 N^{-3}$, which are the decay rates of the most subradiant singly and doubly-excited states. 
$\kappa$ represents the ratio between the two decay rates: $\kappa=\Gamma_{\xi=(1,2)}^{(2)}/\Gamma_{\xi=1}^{(1)}$. 
In Fig.~\ref{fig:kappa}, we plot its numerical value as a function of $N$ and $a/\lambda_{eg}$. In the regime where both modes are subradiant, $\kappa$ is found to be almost constant to a value around $9$.

\section{Justification of the truncation of the Hilbert space to $n_{\textrm{exc}}\leq 2$ in the nonlinear regime}
In this appendix we justify the validity of the numerical truncation of the Hilbert space up to $n_{\textrm{exc}}\leq 2$ for the study of the total population and the instantaneous decay rate.
In Fig.~\ref{fig:3phot}(a,b) we plot the total population in the chain computed with a numerical truncation of the Hilbert space going from $n_{\textrm{exc}}\leq 1$ to $n_{\textrm{exc}}\leq 3$ (in black solid, dashed and pointed lines). Let us denote $\textrm{Pop}_{\textrm{n}_{exc}}$  the total population computed with a numerical truncation up to $\textrm{n}_{exc}$.  We observe that $\textrm{Pop}_{2}$ and $\textrm{Pop}_{3}$ are very similar   during the illumination (a) and the  storage (b) steps of the protocol.
This demonstrates that the numerical truncation to $n_{\textrm{exc}}\leq2$ is enough to properly represent the total population in the system during the different steps of our protocol in the limit of low three-photon population defined as  $(\Omega t_1)^6 \ll 1$.

Let us now  justify the  validity of the numerical truncation to $n_{\textrm{exc}}\leq2$ in the analysis of $t_{\textrm{tr}}$. To do so we plot in Fig.~\ref{fig:3phot}(c) the numerical value of $\gamma(t)$ during the storage computed with a numerical truncation of the Hilbert space from $n_{\textrm{exc}}\leq 1$ to $n_{\textrm{exc}}\leq 3$. 
We see that all curves have the same long time limit (given by the decay rate of the most subradiant singly-excited state). However,  it appears  that the reduction of the analysis to  $n_{\textrm{exc}}\leq 1$ creates a strong error in the short time behavior of the instantaneous decay rate.
 By contrast, comparing  $\gamma(t)$ for $n_{\textrm{exc}}\leq 2$ and $n_{\textrm{exc}}\leq 3$, we  see that the presence of a three-photon population only slightly modifies  the instantaneous decay rate for times shorter than $t_{\textrm{tr}}$.
 From this analysis, we conclude that we can neglect  the tiny three-photon population (and higher manifolds), in the analysis of the $t_{\textrm{tr}}$ made in the nonlinear regime.

\bibliography{bibliomemory}

\end{document}